\documentstyle[preprint,aps,floats,epsfig]{revtex}
\tightenlines
\begin{document}
\title{Elliptic flow in heavy ion collisions near the balance energy}
\bigskip
\author{Yu-Ming Zheng$^{\rm a}$\footnote{
Permanent address: China Institute of Atomic Energy,
P. O. Box 275(18), Beijing 102413, China}, C. M. Ko$^{\rm a}$,
Bao-An Li$^{\rm c}$, and Bin Zhang$^{\rm a}$}
\address{$^{\rm a}$Cyclotron Institute and Physics Department, 
Texas A\&M University\\
College Station, Texas 77843-3366, USA}
\address{$^{\rm c}$Department of Chemistry and Physics, Arkansas State 
University\\
P.O. Box 419, State University, AR 72467-0419, USA}

\maketitle

\begin{abstract} 
The proton elliptic flow in collisions of $^{\rm 48}$Ca on $^{\rm 48}$Ca 
at energies from 30 to 100 MeV/nucleon is studied in an isospin-dependent 
transport model. With increasing incident energy, the elliptic flow shows 
a transition from positive to negative flow.  Its magnitude depends on both 
the nuclear equation of state (EOS) and the nucleon-nucleon scattering cross 
section.  Different elliptic flows are obtained for a stiff EOS with free 
nucleon-nucleon cross sections and a soft EOS with reduced nucleon-nucleon 
cross sections, although both lead to vanishing in-plane transverse flow 
at the same balance energy. The study of both in-plane and elliptic flows 
at intermediate energies thus provides a means to extract simultaneously 
the information on the nuclear equation of state and the nucleon-nucleon 
scattering cross section in medium.
\bigskip

\noindent{PACS number(s): 25.70.-z, 25.75.Ld, 24.10.Lx}
\end{abstract}
\bigskip\bigskip

Heavy ion collisions provide the possibility to study the properties
of nuclear matter in conditions vastly different from that in normal nuclei,
such as high density and excitation as well as large difference in 
the proton and neutron numbers \cite{sto86,cass90,harris96,ko96,bali98}.
Such knowledge is not only of interest in itself but also useful 
in understanding astrophysical phenomena such as the properties of 
the core of compact stars, the evolution of the early universe, and 
the formation of elements in stellar nucleosynthesis. One observable that 
has been extensively used for extracting such information from heavy ion 
collisions is the collective flow of various particles 
\cite{dan85,pan93,zha94,bali93,gqli,wang98,brown,li96,li99,hung95,ris96,sorge,dan98,zhang99,heisel,teney} 
(for a recent review, see Refs. \cite{wes99,awes98,oll98,reis98}). 
For example, the proton flow in heavy ion collisions at 200 MeV/nucleon 
to 1 GeV/nucleon has been found to be consistent with a soft nuclear 
equation of state \cite{pan93,zha94}. From the kaon flow in heavy ion 
collisions at 1 to 2 GeV/nucleon, the existence of a weak repulsive kaon 
potential has been obtained \cite{gqli}.  In heavy ion collisions
at energies higher than 2 GeV/nucleon, recent studies of proton flow 
seem to indicate that there is a softening of nuclear equation of state as 
the nuclear density and excitation increase \cite{dan98}. There are also 
suggestions that particle collective flows at ultrarelativistic heavy ion 
collisions are sensitive to the initial parton dynamics \cite{zhang99} 
and subsequent phase transitions \cite{sorge,heisel,teney}.

In general, collective flow in heavy ion collisions is affected by both 
the nuclear mean-field potential and nucleon-nucleon cross sections. In 
heavy ion collisions at intermediate energies of a few tens MeV/nucleon, 
the collision dynamics is dominated by the attractive nuclear mean-field 
potential as nucleon-nucleon scatterings are largely blocked due to the 
Pauli principle. As a result, the nucleon transverse flow in the reaction 
plane is negative, i.e., nucleons moving in the projectile direction are 
deflected to negative angles. With increasing incident energies, the 
repulsive nucleon-nucleon scattering becomes important and reduces the 
negative flow caused by the attractive nuclear mean-field potential. At 
certain incident energy, called the balance energy, in-plane transverse 
flow vanishes as a result of the cancellation between these two competing 
effects \cite{das93}.  The disappearance of transverse collective flow has 
been experimentally observed in heavy ion collisions \cite{sull,west,ganil}.
The measured balance energy depends strongly on the mass and isospin of 
colliding nuclei as well as on the impact parameter of collisions 
\cite{wes99,awes98,sull,west,ganil}. Studies based on transport models 
have shown that the same balance energy can be obtained with different 
nuclear equations of state and in-medium nucleon-nucleon scattering 
cross sections \cite{kla93,li96a}. To extract their information from 
measured balance energies thus requires the measurement of other 
observables. One of the present authors \cite{li99a} has recently shown 
that different EOS and cross sections which give the same balance energy 
show different differential transverse flows, i.e., their transverse 
flows have different dependence on the total transverse momentum. In 
the present paper, we shall study instead the proton elliptic flow,
which measures the anisotropy in their transverse momentum distribution.
In particular, using the isospin-dependent Boltzmann-Uehling-Uhlenbeck 
(IBUU) model \cite{li96a}, we shall consider collisions of 
$^{\rm 48}$Ca + $^{\rm 48}$Ca at energies from 30 to 100 MeV/nucleon. 
As shown below, different EOS and cross sections that give the
same balance energy lead to significantly different elliptic flows.

Taking the beam direction along the $z-$axis and the reaction plane on 
the $x-z$ plane, the elliptic flow is then determined from the average 
difference between the square of the $x$ and $y$ components of particle 
transverse momentum, i.e.,
\begin{equation}
v_{2} = \left<\frac {p^2_{x} - p^2_{y}}{p^2_{x} + p^2_{y}}\right>.
\end{equation}
It corresponds to the second Fourier coefficient in the transverse momentum 
distribution \cite{oll98,vol97} and describes the eccentricity of an 
ellipse-like distribution, i.e., $v_{2} >0$ indicates in-plane enhancement,
$v_{2} <0$ characterizes the squeeze-out perpendicular to the reaction 
plane, and $v_{2}=0$ shows an isotropic distribution in the transverse plane. 

The IBUU transport used in the present study treats explicitly protons 
and neutrons. It also includes an asymmetry term in the nuclear mean-field 
potential and different scattering cross sections for protons and neutrons. 
The nuclear mean-field potential is parameterized as 
\begin{equation}
U(\rho , \tau _{z}) = U_0(\rho ) + U_{\rm asy}(\rho , \tau _{z}), 
\end{equation}
\begin{equation}
U_{0}(\rho ) = a\left(\frac {\rho}{\rho _{0}}\right) 
+ b\left(\frac {\rho}{\rho _{0}}\right)^{\sigma},
\end{equation}
\begin{equation}
U_{\rm asy}(\rho , \tau _{z}) = C\frac {\rho_{p} - \rho_{n}}{\rho _{0}}\tau_{z}.
\end{equation}
In the above, $\rho _{0}$ is the normal nuclear matter density; $\rho $, 
$\rho_{n}$ and $\rho_{p}$ are the nucleon, neutron, and proton densities, 
respectively; and $\tau _{z}$ equals 1 for proton and -1 for neutron. 
For the strength of the asymmetry potential, we take $C=32$ MeV. Two 
different EOS are used in our studies: a stiff EOS with compressibility 
of 380 MeV ($a=-124$ MeV, $b=70.5$ MeV, $\sigma=2$) and a soft one with 
compressibility of 200 MeV ($a=-356$ MeV, $b=303$ MeV, $\sigma=7/6$). 
We also include the Coulomb potential for protons.  For nucleon-nucleon 
scatterings, both elastic and inelastic channels are included by using 
the experimentally measured cross sections with explicit isospin dependence.
Details of the IBUU model can be found in Refs. \cite{li96a}. 

\bigskip
\begin{figure}[ht]
\centerline{\epsfig{file=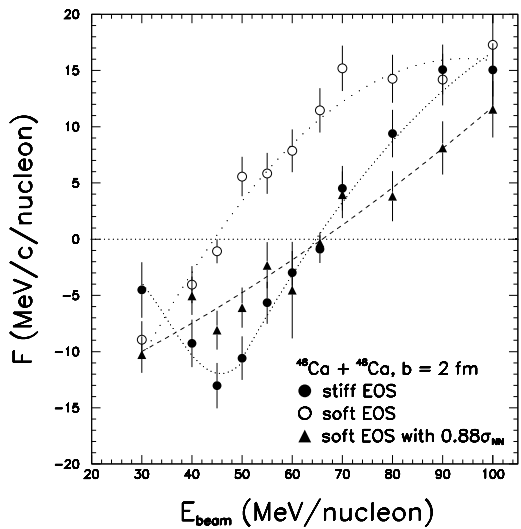,width=3.1in,height=3.1in,angle=0}}
\vspace{0.1in}
\caption{The excitation function of the proton flow parameter in 
$^{\rm 48}$Ca + $^{\rm 48}$Ca collisions at an impact parameter of 2 fm.
Open and solid circles correspond to soft and stiff EOS, respectively.
Solid triangles are from the soft EOS and a reduced nucleon-nucleon 
cross section 0.88$\sigma_{NN}$.}
\label{slope}
\end{figure}

We first study the flow parameter at midrapidity, which is defined by 
\begin{equation}
F={\frac{d<p_x>}{dy}}|_{y=0}.
\end{equation}
In Fig. \ref{slope}, we show the incident energy dependence of the proton 
flow parameter in $^{\rm 48}$Ca + $^{\rm 48}$Ca reactions at an impact 
parameter of 2 fm. Open circles are obtained from the IBUU model using 
the soft EOS. In this case, the proton in-plane transverse flow below
about 45 MeV/nucleon is negative as a result of the dominant effect of 
attractive nuclear mean-field potential. Above this incident energy, 
nucleon-nucleon scatterings become more important, and their repulsive 
effects lead to a positive flow parameter. For the stiff EOS, shown by 
solid circles, the flow parameter is generally reduced because of a less 
attractive mean-field potential than that for the soft EOS.  The exception
to this general behavior occurs, however, at very low incident energies 
below about 40 MeV/nucleon, where the flow parameter increases instead 
with decreasing incident energy. This is due to the fact that scattering 
effects at low energies are not strong enough to reverse the effect due 
to the attractive mean-field potential.  We have also shown in Fig. 
\ref{slope} by solid triangles the flow parameter obtained for the soft 
EOS but with the nucleon-nucleon scattering cross section reduced by 12\%. 
Compared with the case of the soft EOS and free nucleon-nucleon 
cross section $\sigma_{NN}$, the flow parameter is reduced as expected. 
We note that the same balance energy, about 65.5 MeV/nucleon,
is obtained for both the stiff EOS with $\sigma_{NN}$ and the soft EOS with 
$0.88\sigma_{NN}$.

\bigskip
\begin{figure}[ht]
\centerline{\epsfig{file=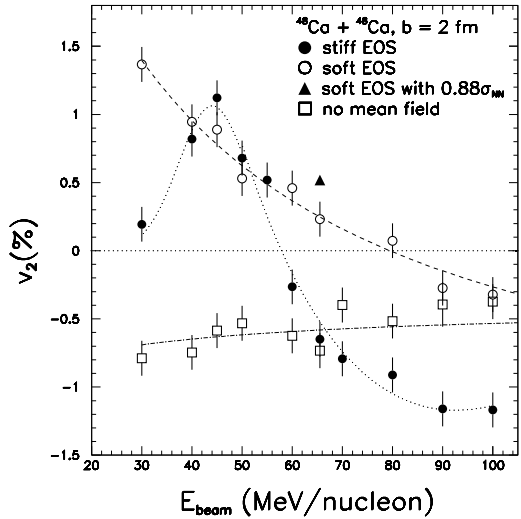,width=3.1in,height=3.1in,angle=0}}
\vspace{0.1in}
\caption{The excitation function of the proton elliptic flow in 
$^{\rm 48}$Ca + $^{\rm 48}$Ca collisions at an impact parameter of 2 fm.}
\label{elliptic}
\end{figure}

The excitation function of the proton elliptic flow for the same reaction is 
shown in Fig. \ref{elliptic}. For both the soft EOS (open circles) and the 
stiff EOS (solid circles) the proton elliptic flow changes from positive 
flow at low energies to negative flow at high energies, i.e., a transition 
from the dominance of in-plane transverse flow to that of out-of-plane 
squeeze out as the beam energy increases. However, the energy at which 
this transition occurs differs for the two EOS; it is smaller for the 
stiff EOS than for the soft EOS. To understand this difference, we also 
show in Fig. \ref{elliptic} the proton elliptic flow in the absence of 
mean-field potential (open squares), which is negative at all energies, 
i.e., out-of-plane squeeze out dominates over in-plane transverse flow. 
Since the soft EOS gives a larger in-plane transverse flow than that due 
to the stiff EOS, it leads to a higher energy at which the elliptic flow 
changes sign. The abnormal behavior at energies below 45 MeV/nucleon, 
where the elliptic flow for the stiff EOS decreases with decreasing
energy, reflects its behavior in the flow parameter as shown in Fig. 
\ref{slope}.

Also shown in Fig. \ref{elliptic} is the elliptic flow for the soft
EOS with 0.88$\sigma_{NN}$ (solid triangle), which gives the same 
flow parameter as the stiff EOS with $\sigma_{NN}$. As seen, the two 
give very different elliptic flow; it is negative for the stiff EOS 
with $\sigma_{NN}$ but is positive for the soft EOS with 0.88$\sigma_{NN}$. 

In summary, the IBUU model has been used to study the proton elliptic 
flow in collisions of $^{\rm 48}$Ca + $^{\rm 48}$Ca at an impact parameter 
of 2 fm for beam energies from 30 to 100 MeV/nucleon.  We find that it shows 
a transition from positive to negative flow as the incident energy increases.  
A strong dependence on both the nuclear EOS and the nucleon-nucleon cross 
section is seen in proton elliptic flow.  Although both the stiff EOS 
with $\sigma_{NN}$ and the soft EOS with $0.88\sigma_{NN}$ have the 
same balance energy, they are found to give very different elliptic flows.
The study of both in-plane and elliptic flows at intermediate energies thus 
allows one to extract simultaneously the information on the nuclear equation
of state and the nucleon-nucleon scattering cross section in medium.

\bigskip
We thank Joe Natowitz for a critical reading of the manuscript.
This work was supported in part by the National Science Foundation 
Grant PHY-9870038, the Department of Energy grant DE-FG03-93ER40773,
the Welch Foundation Grant A-1358, and the Texas Advanced Research 
Program FY97-010366-068.

\end{document}